**Title: Decreased serum vitamin D level as a prognostic marker in patients with COVID-19**

**Ruyi Qu [1#], Qiuji Yang [1#], Yingying Bi [2#], Jiajing Cheng [2], Mengna He [3], Xin Wei [4], Yiqi Yuan [5], Yuxin Yang [6*] and Jinlong Qin[2*]**

[1] Department of Geriatrics, Shanghai Fourth People's Hospital, School of Medicine, Tongji University, Shanghai 200434, China

[2] Department of Obstetrics and Gynecology, Shanghai Fourth People's Hospital, School of Medicine, Tongji University, Shanghai 200434, China

[3] Information Department, Shanghai Fourth People's Hospital, School of Medicine, Tongji University, Shanghai 200434, China

[4] Department of Radiology, Shanghai Fourth People's Hospital, School of Medicine, Tongji University, Shanghai 200434, China

[5] Clinical Laboratory, Shanghai Fourth People's Hospital, School of Medicine, Tongji University, Shanghai 200434, China

[6] Department of Obstetrics and Gynecology, Shanghai Fourth People's Hospital, School of Life and Sciences and Technology, Tongji University, Shanghai 200434, China

* Correspondence: 22310050@tongji.edu.cn (Yuxin Yang); 2180129@tongji.edu.cn (Jinlong Qin)

# These authors contributed equally to this work.

**Abstract**

**Background:** The corona virus disease 2019 (COVID-19) pandemic, which is caused by severe acute respiratory syndrome coronavirus 2, is still localized outbreak and has resulted in a high rate of infection and severe disease in older patients with comorbidities. The vitamin D status of the population has been found to be an important factor that could influence outcome of COVID-19. However, whether vitamin D can lessen the symptoms or severity of COVID-19 still remains controversial.

**Methods:** A total of 719 patients with confirmed COVID-19 were enrolled retrospectively in this study from April 13 to June 6, 2022 at Shanghai Forth People's Hospital. The circulating levels of 25(OH)D3, inflammatory factors, and clinical parameters were assayed. Time to viral RNA clearance (TVRC), classification and prognosis of COVID-19 were used to evaluate the severity of COVID-19 infection.

**Results:** The median age was 76 years (interquartile range, IQR, 64.5-84.6), 44.1% of patients were male, and the TVRC was 11 days (IQR, 7-16) in this population. The median level of 25(OH)D3 was 27.15 (IQR, 19.31-38.89) nmol/L. Patients with lower serum 25(OH)D3 had prolonged time to viral clearance, more obvious inflammatory response, more severe respiratory symptoms and higher risks of impaired hepatic and renal function. Multiple regression analyses revealed that serum 25(OH)D3 level was negatively associated with TVRC independently. ROC curve showed the serum vitamin D level could predict the severity classification and prognosis of COVID-19 significantly.

**Conclusions**: Serum 25(OH)D3 level is independently associated with the severity of COVID-19 in elderly, and it could be used as a predictor of the severity of COVID-19. In addition, supplementation with vitamin D might provide beneficial effects in old patients with COVID-19.

**Keywords:** COVID-19; vitamin D; time to viral RNA clearance; inflammatory response

## Introduction

As of 30 November 2022, the corona virus disease 2019 (COVID-19) pandemic has resulted in more than 639 million confirmed cases with more than 6.6 million deaths[1]. China has also experienced several waves of COVID-19 pandemic and has explored various strategies to protect the susceptible people from infection, especially the elderly and patients with comorbidities. Therefore, it is of great significance to analyze the risk factors of COVID-19 and investigate inventions to reduce the risks of infection or serious illness for the prevention and treatment of COVID-19 in China.

In addition to the injury of alveolar epithelial cells mediated by angiotensin-converting enzyme 2 (ACE2), severe acute respiratory syndrome coronavirus 2 (SARS-CoV-2) could activate macrophages via ACE2 receptors [2, 3]. The interleukins 1 (IL-1), IL-6 and tumor necrosis factor (TNF-α) released by the activated macrophage could further stimulate the neutrophils and T lymphocytes, followed by the release of a large number of inflammatory factors. This inflammatory cascade and inflammation storm are considered to be the main pathogenesis of acute respiratory distress syndrome in COVID-19 [4, 5].

In addition to regulating calcium and phosphorus metabolism, vitamin D is also closely related to immune regulation, cardiovascular diseases, metabolic syndrome, obesity, diabetes, hypertension, cancer, infection and other diseases [6-10]. The multiple effects of vitamin D are related to the wide distribution of the vitamin D receptor (VDR). After binding with VDR, the activated vitamin D acts on the cis-acting elements in the promoter of target genes and thus regulating the transcription of the target genes. Most immune cells, including dendritic cells, T lymphocytes, and B lymphocytes, have high levels of VDR that could modulate the cellular response to viruses as binding with vitamin D [11, 12]. In addition, there are expressions of VDR in the lung tissue, which are related with the severity of lung injury. Previous studies showed that mice with VDR knockout had more serious lung injury induced by LPS compared with WT mice [13, 14]. The histological study showed increased alveolar permeability, pulmonary vascular exudation, neutrophil infiltration and inflammatory factors in the lungs of VDR knockout mice [13, 14].

Although vitamin D has multiple beneficial effects, the nutritional status of vitamin D in the population is unsatisfactory. An epidemiological study in more than 40 countries conducted by Lips P et al. found that vitamin D deficiency was present in more than 50% of the population, especially among nursing home residents (mainly elderly) [15]. In Europe, approximately 40% of the population is vitamin D deficient (< 20 nmol/L), as well as in USA (24%) and Canada (37%) [16, 17].

Previous study showed there was a close relationship between the risk of respiratory tract infection and vitamin D [18]. Patients with daily or weekly vitamin D supplementation, especially those with 25(OH)D3 < 10 nmol/L, were found to have a reduced risk of respiratory tract infections [18]. Studies in patients with COVID-19 also demonstrated that vitamin D levels were related with the infectious risk and severity of COVID-19 [18-20]. A retrospective study by Angelidi et al. found that patients with low serum 25(OH)D3 levels had increased mortality and risk of invasive mechanical ventilation. The median 25(OH)D3 level in this population was 28 nmol/L. The mortality in patients with 25(OH)D3 < 30 nmol/L was 25%, compared with the 9% mortality in patients with > 30 nmol/L. The results were similar when the cutoff value was adjusted as 20 nmol/L [19, 20]. Although treatment with vitamin D failed to improve the survival in critically ill patients with COVID-19 [21], epidemiological studies showed that patients with high vitamin D levels had low risks of infection with COVID-19 [22], which demonstrated that vitamin D could prevent people from COVID-19. Therefore, vitamin D levels have the potential to be a predictor of the risk of infection and the severity of COVID-19. In addition, it is unclear whether some subgroups of patients would benefit from treatment with vitamin D.

In this study, we aimed to assess vitamin D levels in patients with COVID-19 infection, and to investigate the relationship between vitamin D levels and time to clearance of virus, the classification and progression of the disease, which might provide evidence for identifying of high-risk patients for COVID-19, and protecting these vulnerable patients from infection and critical illness of COVID-19.

## Patients and Methods

## Study Population

This is a retrospective cohort study of 719 patients aged 22 to 92 years with confirmed COVID-19 pneumonia hospitalized at the Shanghai Fourth People's Hospital, School of Medicine, Tongji University, Shanghai, China. All patients were diagnosed with COVID-19 pneumonia according to World Health Organization interim guidance. According to hospital data, patients were admitted from April 13 to June 6, 2022, the final date of follow-up was June 20, 2022. The study was approved by the ethics committee of Shanghai Fourth People's Hospital (No. 2020012) and individual consent for this retrospective analysis was waived.

## Data Collection

The epidemiological, clinical evaluation and outcomes data of all participants during hospitalization were collected from electronic medical records by a trained team of physicians. The individual components of clinical outcomes were reviewed independently and recorded into the computer data base by 2 authors (R.L. and Q.L). The clinical outcomes (including the time to viral RNA clearance (TVRC), the classification and progression of COVID-19) were monitored up to June 20, 2022, the final date of follow-up. The Viral Nucleic Acid Kit (Health) was used to extract nucleic acids from clinical throat swab samples obtained from all patients at admission. A 2019-nCoV detection kit (Bioperfectus) was used to detect the ORF1ab gene (nCovORF1ab) and the Ngene (nCoV-NP) according to the manufacturer's instructions using real-time reverse transcriptase–polymerase chain reaction (qPCR). COVID-19 infection was considered laboratory-confirmed if both the nCovORF1ab and nCoV-NP tests showed positive results. Liver and kidney function, lipids and electrolytes were measured by CH930, Atellica Solution, Siemens, Germany. Cytokines were measured by Deflex, Beckman flow cytometry. Blood Routine and C-reactive protein (CRP) were measured by BC7500, Mindray, China. Serum 25 hydroxyvitamin D was determined by cobas 8000, Roche.

## Statistical Analysis

All statistical analyses were performed using IBM SPSS Statistics (Version 22.0, 147 SPSS, IBM Corp., Armonk, New York, USA), GraphPad Prism 8.0.2 (GraphPad

Software, Inc., San Diego CA, USA) and R (version 3.4.1, R Foundation for Statistical Computing, Vienna, Austria). Continuous variables were presented as mean ± Standard Deviation (SD) or median (quartile), and categorical variables were summarized as counts (frequency percentages). $\chi 2$ or Fisher exact test (for small cell counts) was applied to compare categorical variables. For continuous variables, normal distribution was evaluated with Kolmogorov-Smirnov test. Then One-way ANOVA (if homogeneity of variances was assumed) or Wilcoxon-Mann-Whitney U test (if homogeneity of variances was not met) was used. Furthermore, receiver operating characteristics (ROC) curves were performed to investigate the value of serum vitamin D level in predicting the severity classification and prognosis of COVID-19 in the population.

All reported values were two-sided and $P < 0.05$ was considered as statistical significance.

## Results

### 1. Clinical baseline characteristics of enrolled patients

A total of 719 patients with confirmed COVID-19 were enrolled retrospectively in this study from April 13 to June 6, 2022 at Shanghai Forth People's Hospital. In these patients, the median age was 76 years (interquartile range, IQR, 64.5-84.6), 44.1% of patients were male, and the TVRC was 11 days (IQR, 7-16). The median level of 25(OH)D3 was 27.15 (IQR, 19.31-38.89) nmol/L in these patients. The body mass index (BMI) was 23.08 ± 2.59 $Kg/m^2$ in these patients. There were slightly increased levels of mean systolic blood pressure (SBP, 140.85 ± 20.76 mmHg) and respiratory rate (RR, 19.53 ± 1.39 bpm), but normal levels of mean diastolic blood pressure (DBP, 79.85 ± 11.84 mmHg), heart rate (HR, 87.07 ± 15.11 bpm), temperature (Temp, 36.71 ± 0.47℃) and oxygen saturation (SaO2, 97.29 ± 3.54%). The fraction of inspiration O2 (FiO2) was 29 (21- 33) %. The CRP levels were 9.45 (3.39 - 27.9) mg/L, but almost normal levels of white blood cells (WBC, 6.37 ± 3.1×10^9/L), red blood cell (RBC, 4.05 ± 0.70×10^9/L) and hemoglobin (Hb, 121.61 ± 21.07g/L). In addition, the levels of serum bilirubin (Bil), albumin (Alb), alanine aminotransferase (ALT), fasting glucose (FBG) and renal function were in normal ranges (Table 1).

Table 1 Clinical baseline characteristics of enrolled patients.

| Parameter | Value |
|---|---|
| Age (years) | 76.0 (64.5, 84.6) |
| Sex (M/F) | 317/402 |
| TVRC (days) | 11 (7, 16) |
| 25(OH)D3 (nmol/L) | 27.15 (19.31, 38.89) |
| BMI (Kg/m$^2$) | 23.08 ± 2.59 |
| CRP (mg/L) | 9.45 (3.39, 27.9) |
| Temp (℃) | 36.71 ± 0.47 |
| SBP (mmHg) | 140.85 ± 20.76 |
| DBP (mmHg) | 79.85 ± 11.84 |
| HR (bpm) | 87.07 ± 15.11 |
| RR (bpm) | 19.53 ± 1.39 |
| SaO2 (%) | 97.29 ± 3.54 |
| FiO2 (%) | 29 (21, 33) |
| WBC (10^9/L) | 6.37 ± 3.1 |
| RBC (10^12/L) | 4.05 ± 0.70 |
| FBG (mmol/L) | 6.28 ± 2.57 |
| Hb (g/L) | 121.61 ± 21.07 |
| T-Bil (mmol/L) | 13.08 ± 7.89 |
| ALT (U/L) | 19.96 (13.77, 30.87) |
| T-Pro (g/L) | 61.55 ± 6.11 |
| Alb (g/L) | 39.57 (35.81, 42.64) |
| Pre-Alb (g/L) | 183.86 (136.10, 225.50) |
| BUN (mmol/L) | 5.77 (4.57, 7.81) |
| Cr (umol/L) | 57.9 (48.1, 73.7) |
| UA (umol/L) | 288.16 (225.48, 363.44) |
| Cystatin C (mg/mL) | 1.09 (0.91, 1.44) |

Abbreviations: M: male; F: female; TVRC: time to viral RNA clearance; BMI: body mass index; CRP: C-reaction protein; Temp: temperature; SBP: systolic blood pressure; DBP: diastolic blood pressure; HR: heart rate; RR: respiration rate; SaO2: oxygen saturation; FiO2: fraction of inspiration O2; WBC: white blood corpuscle; RBC: red blood corpuscle; FBG: fasting blood glucose; Hb: hemoglobin; T-Bil: total bilirubin; ALT: alanine aminotransferase; T-Pro: total protein; Alb: albumin; Pre-Alb: prealbumin; BUN: blood urea nitrogen; Cr: crea; UA: uric acid.

## 2. Comparison of clinical baseline characteristics and comorbidities among patients with different levels of vitamin D

The levels of serum vitamin D were measured in 609 patients with COVID-19. Then patients were divided into 4 groups according to the quartile values of serum vitamin D levels: Q1 < 13.14 (9.59, 16.56) nmol/L, 13.14 (9.59, 16.56) nmol/L < Q2 < 23.1 (21.37, 25.13) nmol/L, 23.1 (21.37, 25.13) nmol/L < Q3 < 32.42 (29.9, 35.35) nmol/L, and 32.42 (29.9, 35.35) nmol/L < Q4 < 49.29 (43.21, 63.29) nmol/L.

Table 2 showed the clinical baseline characteristics among the four groups. Compared with patients with higher levels of serum vitamin D, patients in Q1 group were older, and had more severe illness, which manifested as longer TVRC, lower oxygen saturation, and FiO2. Patients in the Q1 group also had higher levels of inflammation, which included higher levels of CRP and WBC. There was increased percentage of neutrophil and decreased percentage of monocyte and lymphocyte. Patients in Q1 group also had decreased levels of total protein (T-Pro), Alb, and increased levels of lactate dehydrogenase (LDH), which indicated that patients with low vitamin D levels had impaired hepatic synthetical function and nutritional state. Although there was no significant difference in the blood urea nitrogen (BUN) and Crea (Cr), patients in the Q1 group had increased levels of cystatin C, a biomarker of early renal injury. In addition, there were decreased levels of serum magnesium in patients with lower levels of vitamin D. However, there was no significant difference in male proportion, BMI, basic vital signs, and other biochemical tests (Table 2).

The rates of comorbidities were high in this study. However, there was no significant difference in comorbidities among patients with different levels of vitamin D (Table 2).

Table 2 Comparison of clinical and biochemical characteristics and comorbidities among patients with different levels of vitamin D

| | Q1 group (n=162) | Q2 group (n=164) | Q3 group (n=164) | Q4 group (n=165) |
|---|---|---|---|---|
| Age (years) | 86.0 (70.0, 89.0) | 75.6 (63.75, 87.0)[a] | 72.5 (63.75, 81.25)[ab] | 73.0 (64.0, 81.0)[a] |
| Sex (M/F) | 63/99 | 71/93 | 78/86 | 82/83 |
| TVRC (days) | 14 (8, 19) | 10 (6, 15)[a] | 10 (7.25, 15)[a] | 11 (7, 13)[a] |
| CVD, n (%) | 44 (27.16) | 37 (22.56) | 13 (7.93) | 28 (16.97) |
| HT, n (%) | 98 (60.49) | 84 (51.22) | 93 (56.71) | 89 (53.94) |
| T2DM, n (%) | 36 (22.22) | 36 (21.95) | 48 (29.27) | 50 (30.30) |
| Tumor, n (%) | 17 (10.49) | 15 (9.15) | 15 (9.15) | 14 (8.48) |
| BMI (Kg/m$^2$) | 22.41 ± 3.62 | 22.63 ± 3.57 | 23.68 ± 36.66[ab] | 23.29 ± 3.60 |
| CRP (mg/L) | 15.46 (5.32, 42.08) | 9.45 (3.87, 26.82) [a] | 6.77 (2.68, 19.31) [ab] | 6.29 (2.44, 18.33)[a] |
| Temp (°C) | 36.67 ± 0.46 | 36.66 ± 0.45 | 36.74 ± 0.49 | 36.74 ± 0.49 |
| SBP (mmHg) | 139.61±22.03 | 140.87±22.43 | 141.07±19.76 | 142.95±19.03 |
| DBP (mmHg) | 76.68 ± 12.55 | 81.26 ± 12.43[a] | 80.95 ± 10.80[a] | 80.53 ± 10.90[a] |
| HR (bpm) | 85.41 ± 15.33 | 86.81 ± 17.01 | 89.55 ± 14.5 | 87.24 ± 14.1 |
| RR (bpm) | 19.47 ± 1.63 | 19.66 ± 1.40 | 19.50 ± 1.11 | 19.45 ± 1.15 |
| SaO2 (%) | 96.69 ± 3.40 | 97.46 ± 1.78[a] | 97.52 ± 1.65[a] | 97.62 ± 1.13[a] |
| FiO2 (%) | 29 (29, 33) | 29 (21, 33) [a] | 21 (21, 33) [ab] | 21 (21, 29) [ab] |
| WBC (10^9/L) | 7.14 ± 3.83 | 6.12 ± 2.49[a] | 6.12 ± 2.69[a] | 5.95 ± 3.30[a] |
| Monocyte % | 7.45 ± 2.90 | 8.29 ± 2.70[a] | 8.68 ± 3.31[a] | 8.02 ± 2.96 |
| Lymphocyte % | 21.36 ± 12.13 | 25.60 ± 11.72[a] | 26.58 ± 11.46[a] | 28.76 ± 12.81[ab] |
| Neutrophil % | 69.48 ± 13.30 | 63.84 ± 12.84[a] | 62.37 ± 12.48[a] | 61.10 ± 13.25[a] |
| PLT (10^9/L) | 208.73 ± 90.71 | 211.61 ± 87.55 | 206.95 ± 72.97 | 189.82 ± 68.62[a] |
| RBC (10^12/L) | 3.76 ± 0.76 | 4.10 ± 0.63[a] | 4.19 ± 0.68[a] | 4.20 ± 0.61[a] |
| Hct (%) | 34.19 ± 6.97 | 38.04 ± 5.87[a] | 39.03 ± 5.62[a] | 39.27 ± 5.29[a] |
| Hb (g/L) | 116.36 ± 24.47 | 123.06 ± 19.16[a] | 126.55 ± 17.76[a] | 127.58 ± 17.93[ab] |
| T-Bil (umol/L) | 12.72 ± 7.38 | 13.29 ± 6.60 | 13.38 ± 5.95 | 13.24 ± 10.94 |
| ALT (U/L) | 16.47 (12.77, 28.15) | 22.34 (13.34, 33.24) | 20.14 (13.95, 28.42) | 20.00 (14.57, 32.89) |
| AST (U/L) | 24.83 (19.60, 37.41) | 24.38 (19.23, 32.94) | 23.5 (18.97, 31.12) | 24.59 (19.34, 34.13) |
| AKP (U/L) | 83.41 (67.57, 102.71) | 79.69 (67.88, 99.37) | 78.94 (70.42, 100.54) | 76.97 (61.96, 95.47)[a] |
| T-Pro (g/L) | 58.17 ± 6.46 | 61.74 ± 5.68[a] | 63.20 ± 5.40[ab] | 63.18 ± 5.78[ab] |
| Alb (g/L) | 35.71 ± 4.81 | 39.10 ± 4.33[a] | 40.64 ± 4.16[ab] | 41.02 ± 4.26[ab] |
| Pre-Alb (g/L) | 149.02 (102.26, 203.19) | 179.43 (136.11, 227.65) [a] | 194.49 (162.13, 232.86) [ab] | 190.31 (161, 233.61)[a] |
| BUN (umol/L) | 6.35 (4.68, 9.10) | 5.66 (4.48, 7.16) [a] | 5.77 (4.69, 7.82) | 5.64 (4.57, 7.49) |
| Cr (umol/L) | 56.50 (45.85, 82.90) | 56.3 (48.7, 74.1) | 60.85 (49.35, 72.23) | 59.95 (49.3, 73.55) |
| UA (umol/L) | 251.43 (193.66, 349.31) | 278.97 (239.34, 373.91) | 325.84 (235.18, 372.76) [a] | 295.65 (257.76, 349.96)[a] |
| Cystatin C (mg/mL) | 1.26 (0.98, 1.71) | 1.09 (0.93, 1.38) [a] | 1.05 (0.88, 1.37) [a] | 1.03 (0.89, 1.33)[a] |
| Lactate (mmol/L) | 2.10 ± 0.93 | 1.95 ± 0.76 | 2.04 ± 0.80 | 2.23 ± 1.06 |
| FBG (mmol/L) | 6.63 ± 3.28 | 5.93 ± 2.66[a] | 6.65 ± 3.24[b] | 5.88 ± 2.24[ac] |
| LDH (U/L) | 232.76 ± 84.93 | 209.88 ± 76.38[a] | 203.97 ± 56.08[a] | 201.54 ± 54.01[a] |
| $K^+$ (mmol/L) | 3.90 ± 0.69 | 3.76 ± 0.59[a] | 3.82 ± 0.50 | 3.80 ± 0.50 |
| $Na^+$ (mmol/L) | 141.61 ± 5.77 | 141.94 ± 5.23 | 141.92 ± 3.97 | 142.44 ± 3.55 |
| $Cl^-$ (mmol/L) | 104.11 ± 5.86 | 104.61± 4.83 | 104.36 ± 3.77 | 104.53 ± 3.61 |
| $Ca^{2+}$ (mmol/L) | 1.77 ± 0.46 | 1.87 ± 0.48 | 1.99 ± 0.43 [ab] | 2.05 ± 0.40 [ab] |
| $Mg^{2+}$ (mmol/L) | 0.84 ± 0.11 | 0.87 ± 0.09[a] | 0.88 ± 0.09[a] | 0.87 ± 0.10[a] |
| Phosphate (mmol/L) | 1.08 ± 0.59 | 1.09 ± 0.37 | 1.14 ± 0.23 | 1.15 ± 0.28 |

Abbreviations: CVD: cardiovascular disease; HT: hormone therapy; T2DM: diabetes mellitus type 2; PLT: platelet count; Hct: red blood cell specific volume; AST: glutamic oxaloacetic transaminase; AKP: alkaline phosphatase; LDH: lactate dehydrogenase.

a, $p < 0.05$ compared with Q1 group; b, $p<0.05$ compared with Q2 group; c, $p < 0.05$ compared with Q3 group.

## 3. Comparison of inflammatory factors among patients with different levels of vitamin D

The increased levels of WBC and CRP in patients from Q1 group implicated that patients with lower levels of serum vitamin D might had high inflammatory state. Therefore, we measured the serum levels of inflammatory factors in patients from different groups. However, there was no significant difference of inflammatory factors among these groups except for lower levels of interferon-γ (IFN-γ) and TNF-α in patients with lower levels vitamin D (Figure 1, Table S1).

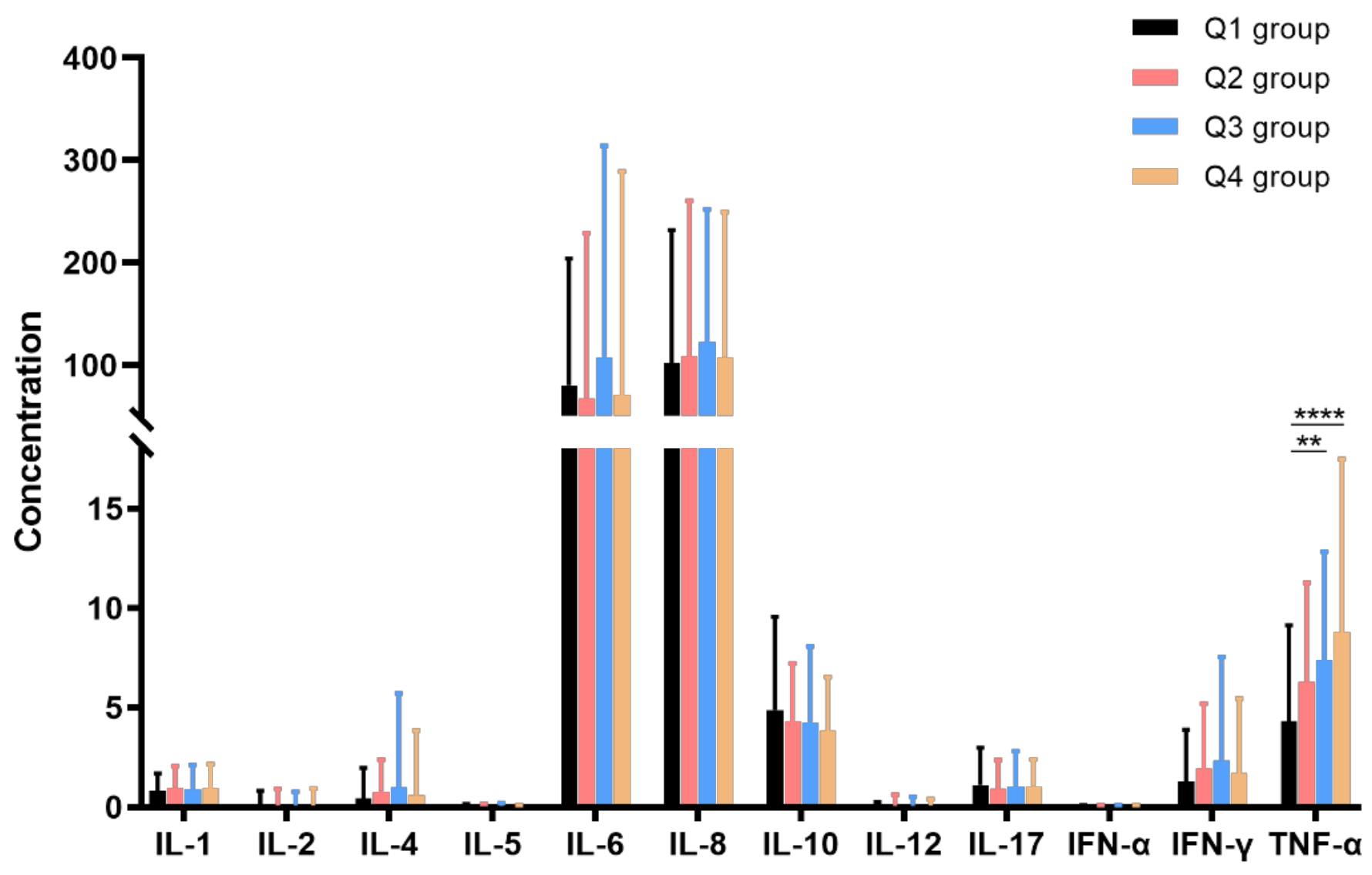

Figure 1 Comparison of inflammatory factors among patients with different levels of vitamin D.**, $p < 0.01$; ****, $p < 0.0001$. Abbreviations: IL: interleukins; IFN: interferon; TNF: tumor necrosis factor.

## 4. Association between serum vitamin D level and the severity of COVID-19

To further assess the association between serum vitamin D level and the severity of COVID-19, patients were first divided into 4 groups according to the quartile of TVRC, which was used as an indicator for the severity of COVID-19. Patients in the longest TVRC group (TVRC-Q4) had significantly lower serum vitamin D levels (23.19 [IQR, 14.46-33.77] nmol/L) compared with patients in shorter TVRC groups (26.74 [IQR, 20.76-38.97] nmol/L in TVRC-Q1, $p = 0.0075$; 31.02 [IQR, 22.87-41.03] nmol/L in TVRC-Q2, $p < 0.0001$; 26.19 [IQR, 18.08, 41.44] nmol/L in TVRC-Q3, $p = 0.0461$) (Figure 2A). Patients were also grouped into mild, moderate, severe and critical groups based on the severity classification of COVID-19 according to the guideline for management of patients with COVID-19 (9th version). There were significantly lower levels of serum vitamin D in patients with severe (19.53 [IQR, 12.71-27.01] nmol/L) and critical (15.54 [IQR, 8.51-20.68] nmol/L) groups compared with patients in the mild (31.10 [IQR, 22.73-42.01] nmol/L) and moderate (26.31 [IQR, 17.98-36.51] nmol/L) groups (Figure 2B). Furthermore, patients were divided into 3 groups based on the prognosis of the disease according to the progression of the disease changes of the severity classification of COVID-19 when the virus RNA was cleared, and the relation between serum vitamin D levels and the prognosis was investigated. Patients with good prognosis had significantly higher levels of serum vitamin D levels (28.21 [IQR, 20.46-40.22] nmol/L) compared with patients with poor prognosis (Prognosis-Q1, 19.53 [IQR, 12.11-27.44] nmol/L in Prognosis-Q2, $p < 0.0001$; 18.03 [IQR, 10.96-21.56] nmol/L in Prognosis-Q3, $p = 0.016$) (Figure 2C).

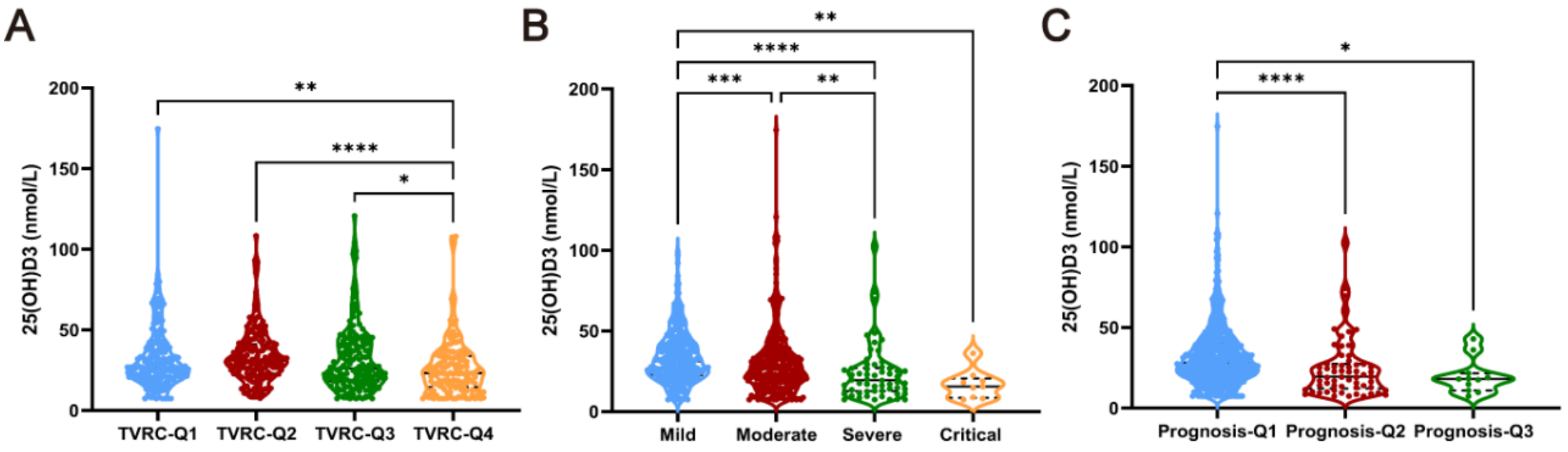

Figure 2 Association of vitamin D level with TVRC, classification and prognosis of COVID-19. **(A)** Vitamin D levels in each group stratified by TVRC quartile 11 (IQR, 7-16). **(B)** Vitamin D levels in each group divided by severity classification of COVID-19. **(C)** Vitamin D levels in patients with different progression.

ROC curve showed the serum vitamin D level could predict the severity classification and prognosis of COVID-19 significantly (the area under the curve [AUC] = 0.695, 95% CI [0.627-0.764], $p < 0.001$, for severe and critical of COVID-19, Figure 3A; AUC=0.728, 95% CI [0.585-0.872], $p = 0.009$, for the aggravation of COVID-19, Figure 3B).

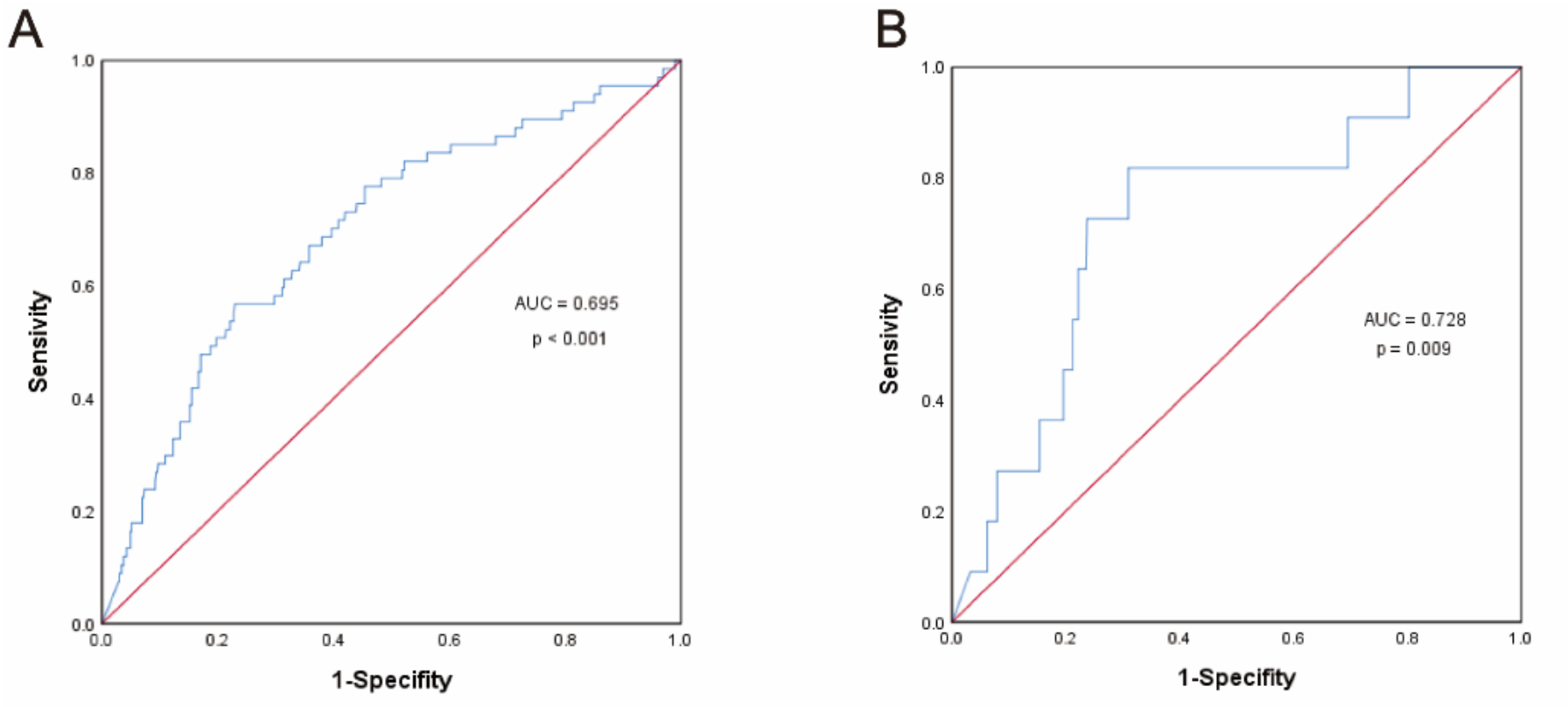

Figure 3 ROC curve to investigate the serum vitamin D level in predicting the severity classification **(A)** and prognosis **(B)** of COVID-19. Abbreviations: AUC: the area under the curve; ROC: receiver operating characteristics.

### 5. Association between serum vitamin D levels and clinical parameters

In univariate analyses, serum vitamin D level was negatively associated with TVRC, age, FiO2, prognosis, IL-10, cystatin C, alkaline phosphatase (AKP), LDH, direct bilirubin (D-Bil), and CRP. However, BMI, SaO2, DBP, Alb, IL-4, TNF-α, serum calcium (Ca) levels, indirect bilirubin (I-Bil), serum magnesium (Mg) level, serum sodium (Na) level, uric acid (UA), pre-albumin (pre-Alb), LDH, Hb, red blood cell specific volume (Hct) and T-Pro were positively associated with serum vitamin D level (Table 3).

Table 3 Correlation between serum vitamin D and other variables

| Parameter | r | p-Value |
|---|---|---|
| Age (years) | -0.239 | < 0.001 |
| TVRC (days) | -0.135 | 0.001 |
| BMI (Kg/m$^2$) | 0.091 | 0.048 |
| CRP (mg/L) | -0.196 | < 0.001 |
| DBP (mmHg) | 0.096 | 0.014 |
| SaO2 (%) | 0.095 | 0.016 |
| FiO2 (%) | -0.227 | < 0.001 |
| WBC (10^9/L) | -0.116 | 0.003 |
| Monocyte % | 0.078 | 0.047 |
| Lymphocyte % | 0.226 | < 0.001 |
| Neutrophil % | -0.23 | < 0.001 |
| Hct (%) | 0.294 | < 0.001 |
| Hb (g/L) | 0.298 | < 0.001 |
| D-Bil (mmol/L) | -0.112 | 0.009 |
| I-Bil (umol/L) | 0.102 | 0.017 |
| AKP (U/L) | -0.104 | 0.035 |
| T-Pro (g/L) | 0.3 | < 0.001 |
| Alb (g/L) | 0.405 | < 0.001 |
| Pre-Alb (g/L) | 0.24 | < 0.001 |
| UA (umol/L) | 0.144 | 0.001 |
| Cystatin C (umol/L) | -0.191 | < 0.001 |
| LDH (U/L) | -0.151 | < 0.001 |
| $Na^{+}$ (mmol/L) | 0.087 | 0.027 |
| $Ca^{2+}$ (mmol/L) | 0.343 | < 0.001 |
| $Mg^{2+}$ (mmol/L) | 0.106 | 0.015 |
| Phosphate (mmol/L) | 0.211 | < 0.001 |
| IL-10 | -0.109 | 0.007 |
| IL-4 | 0.067 | 0.01 |
| TNF-α | 0.202 | < 0.001 |
| DSS | -0.242 | < 0.001 |
| Prognosis | -0.194 | < 0.001 |

Abbreviations: D-Bil: direct Bilirubin; I-Bil: indirect bilirubin; DSS: disease severity score.

**6. Association between TVRC and clinical parameters**

Spearman correlation coefficients were used to evaluate correlations between TVRC and clinical parameters. The results showed that serum vitamin D level, BMI, HR, ALB, TNF-α, serum calcium level, serum sodium level, serum phosphorus level, serum chlorine level, uric acid, pre-Alb, T-Pro, Hb, hematocrit (Hct) and RBC were negatively associated with TVRC. In addition, age, prognosis, IL-10, Il-12, IL-17, IL-2, WBC, CRP, Alb, LDH, ALT, glutamic oxaloacetic transaminase (AST), alkaline phosphatase (AKP), BUN, cystatin C, creatinine, and serum potassium level were positively associated with TVRC (Table 4). Multiple regression analyses revealed that only serum vitamin D level was negatively associated with TVRC independently (Table 5).

Table 4. Correlation between TVRC and other variables

| Variables | Beta coefficient | p-Value |
|---|---|---|
| Age (years) | 0.253 | < 0.001 |
| 25(OH)D3 (nmol/L) | -0.135 | 0.001 |
| BMI (Kg/m$^2$) | -0.164 | < 0.001 |
| CRP (mg/L) | 0.157 | < 0.001 |
| HR (bpm) | -0.074 | 0.047 |
| FiO2 (%) | 0.242 | < 0.001 |
| RBC (10^12/L) | -0.185 | < 0.001 |
| WBC (10^9/L) | 0.09 | 0.016 |
| Lymphocyte % | -0.159 | < 0.001 |
| Neutrophil % | 0.158 | < 0.001 |
| Hct (%) | -0.167 | < 0.001 |
| Hb (g/L) | -0.186 | < 0.001 |
| ALT (U/L) | 0.076 | 0.048 |
| AST (U/L) | 0.081 | 0.03 |
| AKP (U/L) | 0.148 | 0.001 |
| r-GT (U/L) | 0.074 | 0.05 |
| T-Pro (g/L) | -0.167 | < 0.001 |
| Alb (g/L) | -0.287 | < 0.001 |
| Pre-Alb (g/L) | -0.165 | 0.001 |
| BUN (umol/L) | 0.244 | < 0.001 |
| UA (umol/L) | -0.096 | 0.026 |
| Cystatin C (mg/mL) | 0.191 | < 0.001 |
| LDH (U/L) | 0.127 | 0.002 |
| Cr (umol/L) | 0.116 | 0.002 |
| $K^+$ (mmol/L) | 0.207 | < 0.001 |
| $Na^+$ (mmol/L) | -0.204 | < 0.001 |
| $Cl^-$ (mmol/L) | -0.109 | 0.004 |
| $Ca^{2+}$ (mmol/L) | -0.119 | 0.003 |
| Phosphate (mmol/L) | -0.229 | < 0.001 |
| IL-10 | 0.087 | 0.025 |
| IL-12 | 0.08 | 0.04 |
| IL-17 | 0.14 | < 0.001 |
| IL-1 | 0.076 | 0.05 |
| IL-2 | 0.124 | 0.001 |
| TNF-α | -0.095 | 0.014 |
| DSS | 0.235 | < 0.001 |
| Prognosis | 0.178 | < 0.001 |

Abbreviations: r-GT: γ-glutamyl transpeptidase.

Table 5. Multivariate regression analyses of predictors of TVRC in patients with COVID-19

| Variables | Beta coefficient | p-Value | 95% CI |
|---|---|---|---|
| 25(OH)D3 (nmol/L) | -0.230 | 0.016 | -0.168 to -0.018 |

**Discussion**

As a kind of steroid hormone, vitamin D is tightly linked to a number of different metabolic processes and immune regulation in the human body. Vitamin D activates the innate immune system by binding with VDR in immune cells to defend the invasion of foreign pathogenic microorganisms. For example, 1,25 dihydroxyvitamin D3 (1,25-(OH)2-D3) could induce the generation of antimicrobial peptides in monocytes to clean the Mycobacterium tuberculosis[23, 24]. 1,25-(OH)2-D3 also could tune the cellular and humoral immunity by regulating the differentiation and proliferation of T and B lymphocytes and the secretion of Th1/Th2 cytokines. In addition, 1,25-(OH)2-D3 could inhibit the exaggerated inflammatory response via inducing the differentiation of regulatory T cells (Treg), and have protective effects in inflammatory responses and autoimmune diseases.

Considering that 25(OH)D3 is the main form of vitamin D in the body and its stable concentration in circulation, serum 25(OH)D3 was used as an indicator to evaluate the nutritional status of vitamin D. Presently, vitamin D deficiency, insufficiency, normal, and sufficiency are defined as <25, 25 to 50, 51 to 75, and > 75nmol/L, respectively[25]. Vitamin D deficiency was defined when the serum level of 25(OH)D3 was less than 50nmol/L. An epidemiological study in East China showed that the serum levels of 25(OH)D3 were 40.5 ± 12.5 nmol/L in the normal population, and 80.3% of the population were vitamin D deficiency[26], which was significantly higher than that in western countries[27, 28]. In addition, a study in elderly inpatients showed that the vitamin D levels were 34.6 ± 16.2 nmol/L in the population, of which 17.5% were severely deficient, 73.0% were mildly deficient, 7.5% were insufficient, and only 2.0% were sufficient[29]. These data suggest that vitamin D deficiency may be common in the Han population, especially in the elderly and bedridden patients.

This study enrolled 719 patients with COVID-19 and assessed the levels of serum vitamin D, cytokines and other clinical indicators to investigate the relationship between vitamin D levels and TVRC, the classification and prognosis of the disease. Higher levels of vitamin D were associated with the higher levels of T-Pro, Alb, pre-Alb, hemoglobin and BMI, which indicated that higher vitamin D levels were associated with better protein synthesis ability of liver and better nutritional status of patients. Conversely, the lower levels of vitamin D were associated with the longer TVRC, higher levels of WBC and CRP, as well as worse oxygenation capacity of the lung, suggesting that lower vitamin D was associated with severe conditions in these patients. Meanwhile, lower levels of vitamin D were related with the biomarkers of early hepatic and renal function impairments, such as lower levels of pre-Alb and higher levels of cystatin C. In addition, there was positive relationship between vitamin D levels and serum calcium and phosphorus concentrations. All these results demonstrated that vitamin D had benefit effects on the clearance of the virus and alleviating the condition in patients with COVID-19. Further investigation validated that lower vitamin D levels were associated with longer TVRC, more severe disease and worse prognosis. Therefore, serum vitamin D level is a predictor of the severity of disease and prognosis in patients with COVID-19.

Previous studies have shown that the risks of severe infection and mortality were increased in vulnerable groups (with comorbidities such as diabetes, hypertension, coronary artery disease and tumors) in patients with COVID-19[29-32]. In this study, we compared the comorbidities in patients with different vitamin D levels, and found no significant difference in comorbidities among different groups. The results indicated the association between vitamin D levels and the prognosis of the disease was less affected by these chronic comorbidities. Further investigation showed that serum vitamin D level was correlated with TVRC negatively, and serum vitamin D level was an independent predictor of TVRC in patients with COVID-19, which further validated the close relationship between vitamin D and the severity and prognosis of COVID-19.

Vitamin D deficiency is a common phenomenon in Chinese, especially in the elderly. For its detrimental effects on the immune system, vitamin D deficiency would impair the clearance of invasive pathogens. This concern is more obvious under the current situation of the panic of COVID-19 and consistent virus variants. Therefore, it is of great significance to investigate how to protect patients with high risk from infection and improve the prognosis of these patients. Supplement with vitamin D routinely in patients with COVID-19 is still in debate presently[33-35]. However, the results of this study demonstrated that early supplement with vitamin D in patients with COVID-19 and vitamin D deficiency could improve the ability of defensing the infection of SARS-CoV-2, promoting the clearance of virus and improving the prognosis in these high-risk patients. However, for the limitation of the observed study, further prospective randomized controlled trails were needed to investigate the benefits of supplement of vitamin D in these patients.

**Limitations**

There are several limitations of our study. Although our study implied that early supplemented with vitamin D in patients with COVID-19 and vitamin D deficiency might improve the prognosis of these patients. However, we did not give the therapy with vitamin D in this population in this retrospective study. In addition, this retrospective study has some disadvantages compared with prospective studies. Therefore, further prospective studies are needed to validate the clinical value of serum vitamin D levels in risk stratifications of patients with COVID-19.

**Conclusion**

This study demonstrated that serum 25(OH)D3 level was independently associated with the severity of COVID-19 in elderly, and it could be used as a predictor of the severity of COVID-19. In addition, supplementation with vitamin D might provide beneficial effects in old patients with COVID-19.

**Sources of Funding**

This work was supported by Shanghai Committee of Science and Technology, China (grant No. 22dz1202304 to Jiajing Cheng).

**Author Contributions**

Conceptualization, Ruyi Qu, Yuxin Yang and Jinlong Qin; Data curation, Ruyi Qu, Qiuji Yang and Yingying Bi; Formal analysis, Ruyi Qu and Jinlong Qin; Funding acquisition, Jiajing Cheng; Inves-tigation, Jiajing Cheng, Mengna He, Xin Wei and Yiqi Yuan; Methodology, Ruyi Qu, Qiuji Yang, Yingying Bi, Jiajing Cheng, Yuxin

Yang and Jinlong Qin; Project administration, Jinlong Qin; Re-sources, Jiajing Cheng; Software, Yingying Bi and Xin Wei; Supervision, Yuxin Yang and Jinlong Qin; Validation, Qiuji Yang and Yingying Bi; Visualization, Yingying Bi, Mengna He and Yiqi Yuan; Writing – original draft, Ruyi Qu, Qiuji Yang and Yuxin Yang; Writing – review & editing, Yuxin Yang and Jinlong Qin.

**Acknowledgments**

The authors are grateful to all the participants in this study.

**Conflicts of Interest**

The authors declare no conflict of interest.

## Supplemental data

Table 1. Comparison of inflammatory factors among patients with different levels of vitamin D

| | Q1 group (n=150) | Q2 group (n=153) | Q3 group (n=157) | Q4 group (n=149) |
|---|---|---|---|---|
| IL-1 | 0.80(0.16,1.69) | 0.96(0.21,2.03) | 0.89(0.37,2.07) | 0.95(0.29,2.07) |
| IL-2 | 0.08(0.05,0.82) | 0.08(0.04,0.88) | 0.08(0.04,0.77) | 0.08(0.05,0.88) |
| IL-4 | 0.44(0.07,1.90) | 0.78(0.07,2.37) | 0.98(0.08,5.47) | 0.6(0.08,3.77) |
| IL-5 | 0.07(0.03,0.14) | 0.07(0.04,0.14) | 0.07(0.03,0.21) | 0.07(0.04,0.1) |
| IL-6 | 80.44(21.22,204.04) | 67.63(24.65,228.57) | 107.85(26.4,312.4) | 70.84(19.37,288.68) |
| IL-8 | 101.77(32.25,229.92) | 108.2(23.96,255.95) | 122.3(44.53,242.43) | 108.06(32.42,244.3) |
| IL-10 | 4.88(2.79,9.32) | 4.28(2.51,7.21) | 4.26(2.53,8.01) | 3.85(2.09,6.45)[a] |
| IL-12 | 0.07(0.04,0.18) | 0.07(0.04,0.62) | 0.07(0.04,0.5) | 0.08(0.04,0.4) |
| IL-17 | 1.11(0.36,2.93) | 0.91(0.26,2.37) | 1.05(0.33,2.67) | 1.05(0.32,2.31) |
| IFN-α | 0.07(0.04,0.10) | 0.07(0.03,0.09) | 0.07(0.04,0.1) | 0.06(0.04,0.1) |
| IFN-γ | 1.27(0.38,3.79) | 1.93(0.56,5)[a] | 2.35(0.4,7.37)[a] | 1.72(0.42,5.34) |
| TNF-α | 4.32(1.82,9.12) | 6.29(2.91,11.04)[a] | 7.38(3.52,12.74)[a] | 8.78(3.39,17.19)[ab] |

Table 2. Comparison of clinical and biochemical characteristics among patients with different severity rating

| | Q1 group (n=330) | Q2 group (n=309) | Q3 group (n=71) | Q4 group (n=9) |
|---|---|---|---|---|
| Age (years) | 68(59,78) | 81(71,88) [a] | 86(77,90) [a] | 87(79.5,91) [a] |
| Sex (M/F) | 149/181 | 130/179 | 34/37 | 4/5 |
| TVRC (days) | 9(6,13) | 12(8,17) [a] | 13(9,20.25) [a] | 12(3.25,14.75) |
| 25(OH)D3 | 31.10(22.73,42.01) | 26.31(17.98,36.51) [a] | 19.53(12.71,27.01) [ab] | 15.54(8.51,20.68) [ab] |
| BMI (Kg/m$^2$) | 23.42±3.59 | 22.74±3.61 | 21.73±3.18 | NA |
| CRP (mg/L) | 5.46(2.33,14.47) | 10.27(4.38,27.91) [a] | 36.65(19.13,76.31) [ab] | 99.08(56.11,155.47) [ab] |
| Temp (℃) | 36.74±0.48 | 36.68±0.46 | 36.70±0.50 | 37.00±0.41 |
| SBP (mmHg) | 141.06±19.57 | 140.53±21.84 | 141.38±21.72 | 140.00±21.47 |
| DBP (mmHg) | 81.42±11.06 | 78.75±12.02 | 77.87±13.31 | 76.11±15.60 |
| HR (bpm) | 88.33±15.10 | 86.01±14.55 | 85.58±17.10 | 89.22±17.14 |
| RR (bpm) | 19.41±1.24 | 19.58±1.23 | 19.69±2.14 | 20.78±2.99 |
| SaO2 (%) | 97.72±1.21 | 97.38±1.49 | 95.57±3.22 [ab] | 91.38±11.88 [abc] |
| Fio2 (%) | 21(21,29) | 29(21, 33) [a] | 33(33,41) [ab] | 61(29,141) [ab] |
| RBC (10^12/L) | 4.25±0.57 | 3.99±0.72 [a] | 3.53±0.74 [ab] | 3.20±0.71 [abc] |
| WBC (10^9/L) | 5.75±2.58 | 6.45±3.03 [a] | 8.41±3.93 [ab] | 10.49±5.51 [abc] |
| Monocyte % | 8.27±2.94 | 8.15±2.92 | 6.54±2.75 | 7.97±6.25 |
| Lymphocyte % | 29.23±11.37 | 24.32±11.64 [a] | 13.59±8.56 [ab] | 8.78±5.00 [abc] |
| Neutrophil % | 60.23±11.98 | 65.46±12.49 [a] | 78.03±10.61 [ab] | 82.88±9.49 [abc] |
| PLT (10^9/L) | 196.96±65.99 | 210.42±86.97 | 221.68±94.84 | 252.33±159.25 |
| Hct (%) | 39.35±5.35 | 36.91±6.27 [a] | 32.64±6.81 [ab] | 29.43±6.56 [abc] |
| Hb (g/L) | 127.60±18.05 | 118.93±20.43 [a] | 107.37±26.52 [ab] | 106.56±21.07 [ab] |
| T-Bil (umol/L) | 12.73±5.94 | 12.89±6.22 | 14.98±16.83 | 17.20±9.87 |
| ALT (U/L) | 19.93(13.52,30.35) | 20.20(13.91,31.77) | 20.35(13.17,30.83) | 19.02(14.79,30.27) |
| AST (U/L) | 22.71(18.48,29.55) | 24.98(19.32,34.96) [a] | 30.38(21.24,42.76) [a] | 45.24(34.48,50.38) [a] |
| AKP (U/L) | 78.10(63.65,95.12) | 83.77(69.13,99.65) | 80.64(67.78,102.90) | 84.03(67.08,112.51) |
| T-Pro (g/L) | 62.91±5.54 | 61.30±5.92 | 57.54±6.56 [ab] | 52.45±7.15 [abc] |
| Alb (g/L) | 41.01±4.28 | 38.29±4.43 [a] | 34.38±4.45 [ab] | 31.75±4.24 [abc] |
| Pre-Alb (g/L) | 196.58(160.86,238.85) | 181.60(127.98,291.55) [a] | 98.48(80.37,148.75) [ab] | 98.73(65.29,151.81) [a] |
| BUN (umol/L) | 5.43(4.39,6.79) | 6.13(4.80,8.32) [a] | 6.33(4.92,11.15) [a] | 13.85(7.56,20.02) [a] |
| Cr (umol/L) | 57.50(48.80,71.28) | 58.40(47.80,78.70) | 54.20(37.70,80.60) | 89.50(38.10,169.30) |
| UA (umol/L) | 299.76(243.44,359.89) | 291.43(224.63,376.58) | 242.34(142.73,319.98) [ab] | 252.97(148.49,377.04) [a] |
| Cystatin C (mg/mL) | 0.98(0.86,1.21) | 1.16(0.95,1.56) [a] | 1.39(1.06,1.88) [ab] | 1.60(1.16,2.55) [a] |
| Lactate (mmol/L) | 1.94±0.71 | 2.00±0.78 | 2.21±1.04 | 2.77±1.23 |
| FBG (mmol/L) | 5.81±2.72 | 6.35±2.88 [a] | 7.69±3.00 [ab] | 8.07±1.25 [abc] |
| LDH (U/L) | 194.28±48.28 | 215.18±70.80 [a] | 244.00±90.07 [ab] | 358.20±107.77 [abc] |
| K (mmol/L) | 3.75±0.51 | 3.87±0.63 | 3.89±0.54 | 4.20±0.66 |
| Na(mmol/L) | 142.95±3.54 | 141.19±5.27 | 140.75±6.28 | 139.56±7.09 |
| Cl (mmol/L) | 105.00±3.52 | 103.94±5.25 | 103.75±6.22 | 102.33±7.35 [ab] |
| Ca (mmol/L) | 2.05±0.37 | 1.84±0.49 [a] | 1.53±0.49 [ab] | 1.68±0.44 [ab] |
| Mg (mmol/L) | 0.88±0.08 | 0.86±0.10 | 0.82±0.11 [a] | 0.81±0.10 [a] |
| P (mmol/L) | 1.19±0.34 | 1.12±0.43 | 0.88±0.36 [ab] | 0.72±0.37 [abc] |

Table 3. Comparison of clinical and biochemical characteristics among patients with different prognosis

| | P1 group (n=638) | P2 group (n=70) | P3 group (n=11) |
|---|---|---|---|
| Age (years) | 74.5(64,86) | 85.5(78.75,90.25) [a] | 81(66,89) |
| Sex (M/F) | 277/361 | 34/36 | 6/5 |
| TVRC (days) | 10(7,15) | 14(9,23) [a] | 18.5(14.75,22) [a] |
| BMI ($Kg/m^2$) | 23.17±3.60 | 21.79±2.70 | 19.92±4.98 |
| 25(OH)D3 | 28.21(20.46,40.22) | 19.53(12.11,27.44) [a] | 18.03(10.96,21.56) [a] |
| CRP (mg/L) | 7.22(2.96,20.25) | 45.98(22.03,93.56) [a] | 69.08(17.99,105.47) [a] |
| Temp (℃) | 36.71±0.47 | 36.71±0.49 | 36.78±0.41 |
| SBP (mmHg) | 140.69±20.61 | 142.04±22.37 | 142.55±20.86 |
| DBP (mmHg) | 79.84±11.47 | 80.24±14.89 | 77.91±12.76 |
| HR (bpm) | 87.47±14.69 | 83.21±17.82 | 88.45±18.97 |
| RR (bpm) | 19.48±1.22 | 19.99±2.36 [a] | 19.64±1.86 |
| SaO2 (%) | 97.49±1.52 | 95.68±4.87 [a] | 96.00±3.55 |
| Fio2 (%) | 29(21,33) | 33(29, 41) [a] | 41(37,53) [a] |
| RBC (10^12/L) | 4.12±0.65 | 3.47±0.80 [a] | 3.72±0.64 |
| WBC (10^9/L) | 6.09±2.84 | 8.35±3.54 [a] | 10.12±6.57 [a] |
| Monocyte % | 8.21±2.94 | 6.95±2.83 [a] | 5.48±5.14 [a] |
| Lymphocyte % | 26.83±11.81 | 13.69±7.76 [a] | 11.76±11.01 [a] |
| Neutrophil % | 62.76±12.58 | 77.69±9.44 [a] | 82.25±13.39 [a] |
| PLT (10^9/L) | 205.06±78.08 | 211.16±93.12 | 219.82±133.25 |
| Hct (%) | 38.16±5.90 | 32.22±7.32 [a] | 33.75±6.11 |
| Hb (g/L) | 123.41±19.53 | 104.79±23.48 [a] | 125.00±39.87 [b] |
| T-Bil (umol/L) | 12.80±6.08 | 15.83±17.02 [a] | 11.66±6.80 |
| ALT (U/L) | 20.08(13.79,30.75) | 18.68(13.30,32.71) [a] | 21.80(10.16,32.22) [a] |
| AST (U/L) | 23.58(18.86,32.43) | 31.92(20.11,47.30) [a] | 35.85(23.21,47.03) |
| AKP (U/L) | 79.99(65.62,95.93) | 88.07(69.46,113.97) | 80.64(65.92,98.15) |
| T-Pro (g/L) | 62.21±5.75 | 56.77±6.47 [a] | 54.37±6.76 [a] |
| Alb (g/L) | 39.72±4.53 | 34.15±4.47 [a] | 32.74±4.76 [a] |
| Pre-Alb (g/L) | 190.22(147.24,232.20) | 100.75(81.27,146.45) [a] | 129.31(82.19,197.08) |
| BUN (umol/L) | 5.65(4.51,7.43) | 7.32(5.28,12.94) [a] | 10.41(6.20,18.20) [a] |
| Cr (umol/L) | 57.80(48.20,72.48) | 62.15(41.40,90.65) | 67.20(32.10,129.40) |
| UA (umol/L) | 291.73(230.62,365.03) | 275.16(171.31,363.57) | 148.57(79.31,240.75) [ab] |
| Cystatin C (mg/mL) | 1.05(0.89,1.38) | 1.47(1.23,2.15) [a] | 1.20(1.01,2.35) |
| Lactate (mmol/L) | 1.96±0.74 | 2.24±1.07 | 2.89±1.09 [a] |
| FBG (mmol/L) | 6.00±2.67 | 8.07±3.41 [a] | 9.64±3.43 [a] |
| LDH (U/L) | 203.50±58.94 | 251.11±97.54 [a] | 334.63±115.47 [ab] |
| K (mmol/L) | 3.79±0.55 | 4.01±0.71 [a] | 4.17±0.62 |
| Na(mmol/L) | 142.20±4.30 | 139.64±6.25 [a] | 140.82±12.58 |
| Cl (mmol/L) | 104.46±4.39 | 103.93±5.77 | 102.91±11.60 |
| Ca (mmol/L) | 1.93±0.45 | 1.67±0.50 [a] | 1.29±0.39 [ab] |
| Mg (mmol/L) | 0.87±0.09 | 0.83±0.11 [a] | 0.81±0.10 |
| P (mmol/L) | 1.15±0.39 | 0.95±0.37 [a] | 0.63±0.19 [ab] |